\documentclass[showpacs,twocolumn,preprintnumbers,amsmath,amssymb] {revtex4}
\usepackage{graphicx,color,epsfig}
\usepackage{dcolumn}
\usepackage{bm}

\begin{document}
\topmargin 0.2cm

\title{ \bf Three loop correction in the formation of QGP droplet}

\author{M. Jena, K. K. Gupta$^1$ and S. Somorendro Singh\footnote{email:sssingh@physics.du.ac.in} }
\affiliation {Department of Physics and Astrophysics, University of Delhi, Delhi - 110007, India \\
$^1$ Department of Physics, Ramjas College, University of Delhi, Delhi-110007, India \\
}        
%\maketitle

%\begin{center}
%Department of Physics and Astrophysics, University of Delhi, Delhi - 110007, India \\
%\end{center}
%\linadj{200} 
\begin{abstract}
Quark-gluon plasma (QGP) droplet formation
is re-considered with the addition of
three loop correction to the earlier loop factors 
in the mean field potential. The correction of the three loop factor
increases stability in the droplet formations of QGP at different 
parametrization factors of the QGP fluid and it is in better agreement 
in comparison to the lattice results of pressure, energy density and other
thermodynamic relations. This implies that the contribution of the three loop
enhances in showing the characteristic features of the QGP fluid. It shows 
that increasing the loop increased the strength of parametrization value 
which we defined earlier as a number parameter of fluid dynamics. 
It indicates that the model with the loop correction boosts in 
explaining about the formation of QGP droplet in the expansion of 
early universe. 
 
\end{abstract}
%\vfill
%\eject
\maketitle

\section{Introduction}

There are number of reports about the process of phase transition that
occured during the earliy universe formation. These reports are available 
when we study the literature of early universe evolution.
In the early universe, it is believed that the matter about a few microseconds
is made of free quarks
and gluons and subsequently cool down over time, 
the matter behave as confined matter
of bound quarks of hadrons~\cite{jc,dav}. This short period 
life of free quarks and 
gluons is known as quark-gluon plasma (QGP). During the short period of time, 
the universe is highly complicated and it is very difficult to predict the exact
scenarios of the universe. So many theorists and experimentalists try to do 
something about the detection of this complicated matter. 
Although studying the matter is old enough to say it has not been possible
to get whole
information about the critical point and the matter at high baryon 
density. To explain the system, a  
number of highly energetic laboratories have been set up around the globe 
to find out how the early universe was created~\cite{fu,sad}. Due to these experimental facilities available in the different 
places, the attraction of more expertise peoples are increasing day by day
to search about the matter and really become an exciting field in the 
present day of heavy ion collider physics~\cite{satz,ka1,ka2}. 
So heavy-ion-collision is providing the update information about the QGP
formation from time to time and on the basis of its reports, many scientists
around the world theoretically as well as experimentally are formalising 
different modelings to prove the existence and characterization of the
deconfined phase of QGP. Among these modelings we also attempt to develop 
a simple model to represent the structure of QCD and phase transition phenomena.
The model is based on the loop correction factor by creating a mean field
potential among the participating particles in the system. The loop correction
has already shown some details about the thermodynamical pictures of QGP and 
now we extend our earlier work~\cite{s1} by adding the three loop factor 
and looking for the improvement produced by this three loop. 
The model creates the free energies through the different
quark and gluon flow parameters forming various
sizes of stable droplets. The formation of these stable droplet differs
with the change of temperature and parametrization
values. It indicates that 
droplet formation is dependent on both the parametrization value of quark 
and gluon
and the temperature. In addition to it, the critical radius
of the droplet helps in determining the surface tension as it carries 
an important
property of liquid drop model in determining the stability 
of droplet formation in the system.
\par In this paper, we focus on the
QGP droplet formation and its thermodynamical properties through the 
incorporation of three loop correction in mean field potential. This
is however an extensive work of our earlier programs of 
loop corrections~\cite{s1,s2}. Due to
the addition of three loop term to potential
the results produced are successfully able to explain the relevant 
properties of QGP droplets.
\par The paper is arranged as follows: In section $II$, we briefly
construct the Hamiltonian of the system
incorporating three loop correction connecting from the two 
loop correction factor in the potential and find the density of states of 
the free energies. 
In section $III$ the free energy, thermodynamic relations and
surface tension of the system symbolising the stable
droplet formation of QGP are explained. In section $IV$, the analytical
and numerical solutions as results are discussed. 
At last, a brief conclusion with the
details of QGP formation are presented.

\section{Hamiltonian and density of states through loop corrections}
 The interacting potential created among quark-quark,quark-antiquark, 
quarks and gluons is used as mean field potential, which in turn, 
help us to find the density of states of the free energy.
The mean field potential in perturbed QCD is normally used among 
the massive quarks. The same technique of finding the mean field potential 
is applied among the mass-less 
quarks as the mass is considered as quasi-model of temperature dependent
quark mass. It is because of the fact that the system is involving at
very high temperature and its effect is not negligible at all in the
system. There is 
hence contribution of interacting potential in the mean field potential
due to the temperature dependent quark mass.
Therefore from one loop
to three loop correction among the internal quarks are
required for the calculation of the actual interacting potential.
The contributions of these terms are also already reported earlier by our 
works.
So we use the effective mean field 
potential produced by all these loop correction factors in determining
the density of states similar to our earlier papers~\cite{somoren,sss}. 
Due to three loop correction
we get an improved results in all the thermodynamic
parameters. Now Hamiltonian of the confining/de-confining potentials 
among the particles is formulated with the addition of three loop
to the earlier loops and hence it leads
in obtaining the thermal mass of the quarks as~\cite{peshier,gol,rama}:
\begin{eqnarray}
H(k,T) &=&[k^{2}+m^{2}(T)]^{1/2} \nonumber\\
       &=& k+m^{2}(T)/2k~~for~large~k  
\end{eqnarray}
where ,
\begin{eqnarray} 
m^{2}(T)&=&16 \pi \gamma_{qg} ~ \alpha_{s}(k) T^{2} [1+\frac{\alpha_{s}(k)}{4\pi} a_{1} \nonumber \\
&+&\frac{\alpha_{s}^{2}(k)}{16 \pi^2}a_{2} 
+\frac{\alpha_{s}^3(k)}{64 \pi^3}a_{3}].
\end{eqnarray} 
where $\gamma_{qg}=\frac{\sqrt(\gamma_{q}^2+\gamma_{g}^2)}{\gamma_{q}\gamma_{g}}$with $\gamma_{q}=1/10$ and $\gamma_{g}=60\gamma_{q}$. The mass is called 
thermal mass and it is obtained after
the three loop corrections being introduced in the potential.
The corresponding co-efficients obtained in thermal mass are $a_{1}$, $a_{2}$ 
and $a_{3}$.
These are obtained by the interactions among
the constituent
particles through the loop corrections. These are numerically 
found to be depending on the number
of quark flavours and they are given as:
\begin{eqnarray}
a_{1}&=&2.5833-0.2778~ n_{l}, \\
a_{2}&=&28.5468-4.1471~n_{l}+0.0772~n_{l}^{2}, \\
a_{3}&=&209.884-51.4048~n_{l}+2.9061~n_{l}^2-0.0214~n_{l}^3
\end{eqnarray}
where $n_{l}$ is considered to be the number of light quark 
elements~\cite{fischler,bil,smirnov,smi}.
$k$ is the quark (gluon) momentum.  The loop co-efficients $a_{1}$,$a_{2}$ 
and $a_{3}$ play the roles for involving in the creation
of QGP energy and formation of QGP droplet.
Due to these co-efficients, we found the different quark and gluon 
parametrization factors in forming the QGP droplets. The parametrization 
factors are obtained as $ \gamma_{q}=1/14 $ and 
$\gamma_{g}=~ (60 - 70)~ \gamma_{q}$ in obtaining the stable droplets.
It indicates that the overall calculation is controlled with the most
favourable results in all properties. The value varies
from small to large depending on non-loop
to loop correction.
$\alpha_{s}(k)$ is QCD running coupling constant defined as:
\begin{equation}
 \alpha_{s}(k)=\frac{4 \pi}{(33-2n_{f})\ln(1+k^{2}/\Lambda^{2})},
\end{equation}
in which $~\Lambda $ is QCD parameter almost equal to~$ 0.15~$GeV. 
$n_{f}$ is degree of freedom of quark and gluon. 
So the interacting mean-field potential $ V_{conf}(k) $ is
derived with inclusion of one to three loop corrections.
In the derivation of the potential, 
the expansion of strong coupling constant through the one to three
loop factors is used with the technique of
the perturbation theory~\cite{ramanathan,brambilla,melnikov,ho,va}.
\begin{eqnarray}\label{3.18}
V_{\mbox{conf}}(k) &=& \frac{8 \pi}{k}\gamma_{qg} ~ \alpha_{s}(k) T^{2} [1+\frac{\alpha_{s}(k)a_{1}}{4\pi}  
+\frac{\alpha_{s}^2(k)a_{2}}{16 \pi^2} \nonumber \\
&+&\frac{\alpha_{s}^3(k)}{64 \pi^3}a_{3}] - \frac{m_{0}^{2}}{2 k},
\end{eqnarray}

Now the density of states 
in phase space with loop corrections
in the interacting potential is obtained as~\cite{rama,linde,fe,th,be}: 
\begin{equation}
 \rho_{q,g} (k) = \frac{v^2}{3 \pi^{2}}\frac{dV^{3}_{conf}(k)}{dk}~,
\end{equation}
or,
\begin{equation}\label{3.13} 
\rho_{q, g}(k) =\frac{v^2g^6(k)}{8 \pi^2 k^4}\gamma_{qg}^3T^6[1+\frac{\alpha_{s}(k)a_{1}}{\pi}+\frac{\alpha^2(k) a_{2}}{\pi^2}+\frac{\alpha^3(k) a_{3}}{\pi^3}]^2B, 
\end{equation}
where
\begin{eqnarray}
B&=&[\{ 1+\frac{\alpha_{s}(k)a_{1}}{\pi}
+\frac{\alpha_{s}^2(k)a_{2}}{\pi^2} 
+\frac{\alpha_{s}^3(k)a_{3}}{\pi^3}\} \nonumber \\
&+& \frac{2 k^2}{\Lambda^2+k^2}\{\frac{(1+2\alpha_{s}(k)a_{1}/\pi)}{\ln(1+k^2/\Lambda^2)}+\frac{3\alpha_{s}^2(k)a_{2}}{\pi^2} \nonumber \\
&+&\frac{4\alpha_{s}^3(k)}{\pi^3}\frac{a_{3}}{\ln(1+k^2/\Lambda^2)}\}] 
\end{eqnarray}
and~ $v$ is the volume occupied by the QGP and
$~g^{2}(k)=4 \pi \alpha_{s}(k)$. 
%\begin{figure}[h]
%\centering
%\sidecaption
%\includegraphics[width=7cm,clip]{fig1.eps}
%\caption{ free energy contribution~$F_{i}$~vs. R at~$\gamma_{q}=1/14~$, $\gamma_{g}=44\gamma_{q}$
%at the different temperatures~}
%\label{fig-1}       % Give a unique label
%\end{figure}

\begin{figure*}[htb]
\centering
%\sidecaption
\includegraphics[width=7cm,clip]{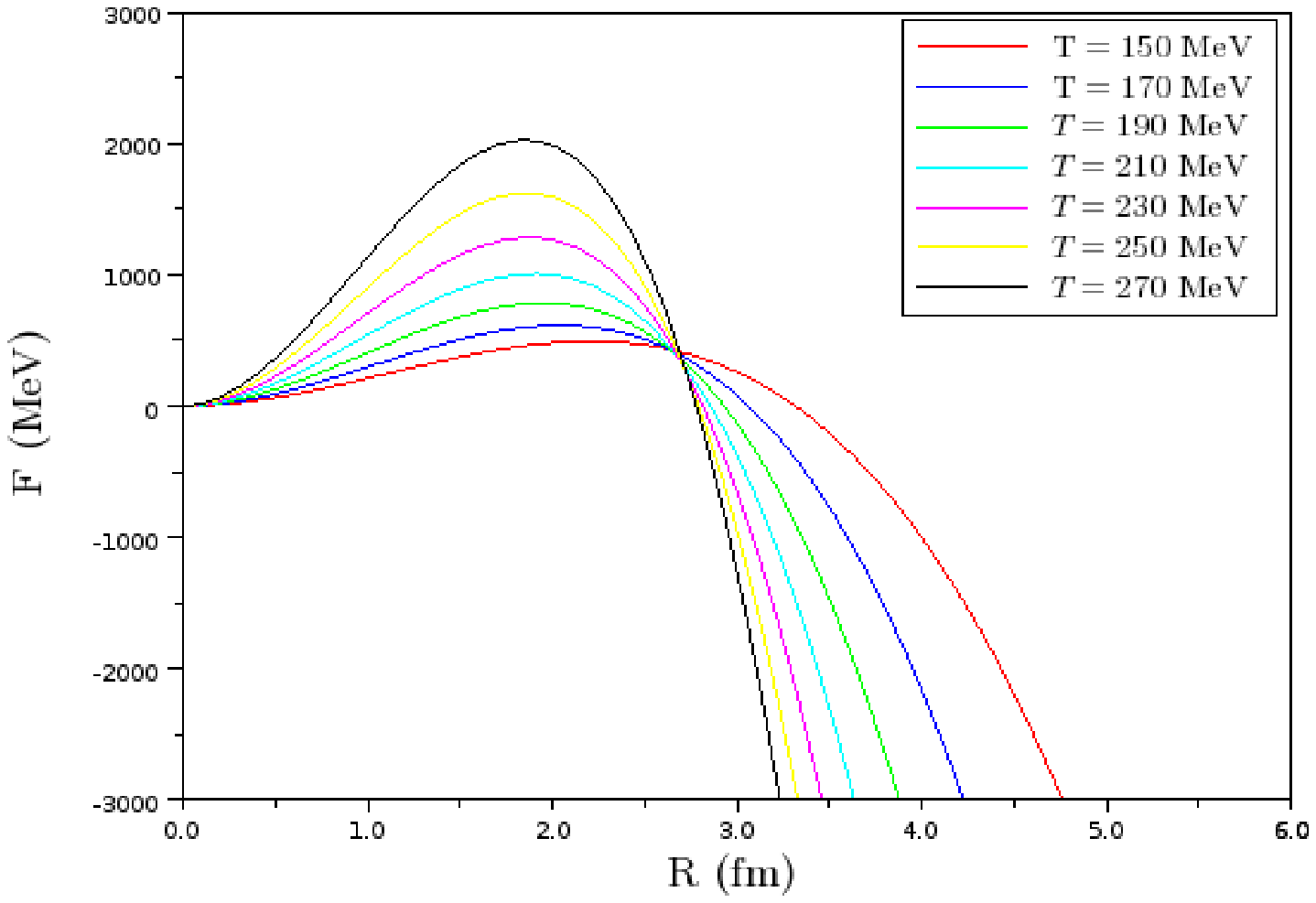}
%\vspace*{5cm}       % Give the correct figure height in cm
\caption{The free energy vs.~R~at~$\gamma_{q}=1/14~$, $\gamma_{g}=60\gamma_{q}$ for various values of temperature.}
\label{fig-2}       % Give a unique label
\end{figure*}

\begin{figure}[htb]
\centering
%\sidecaption
\includegraphics[width=7cm,clip]{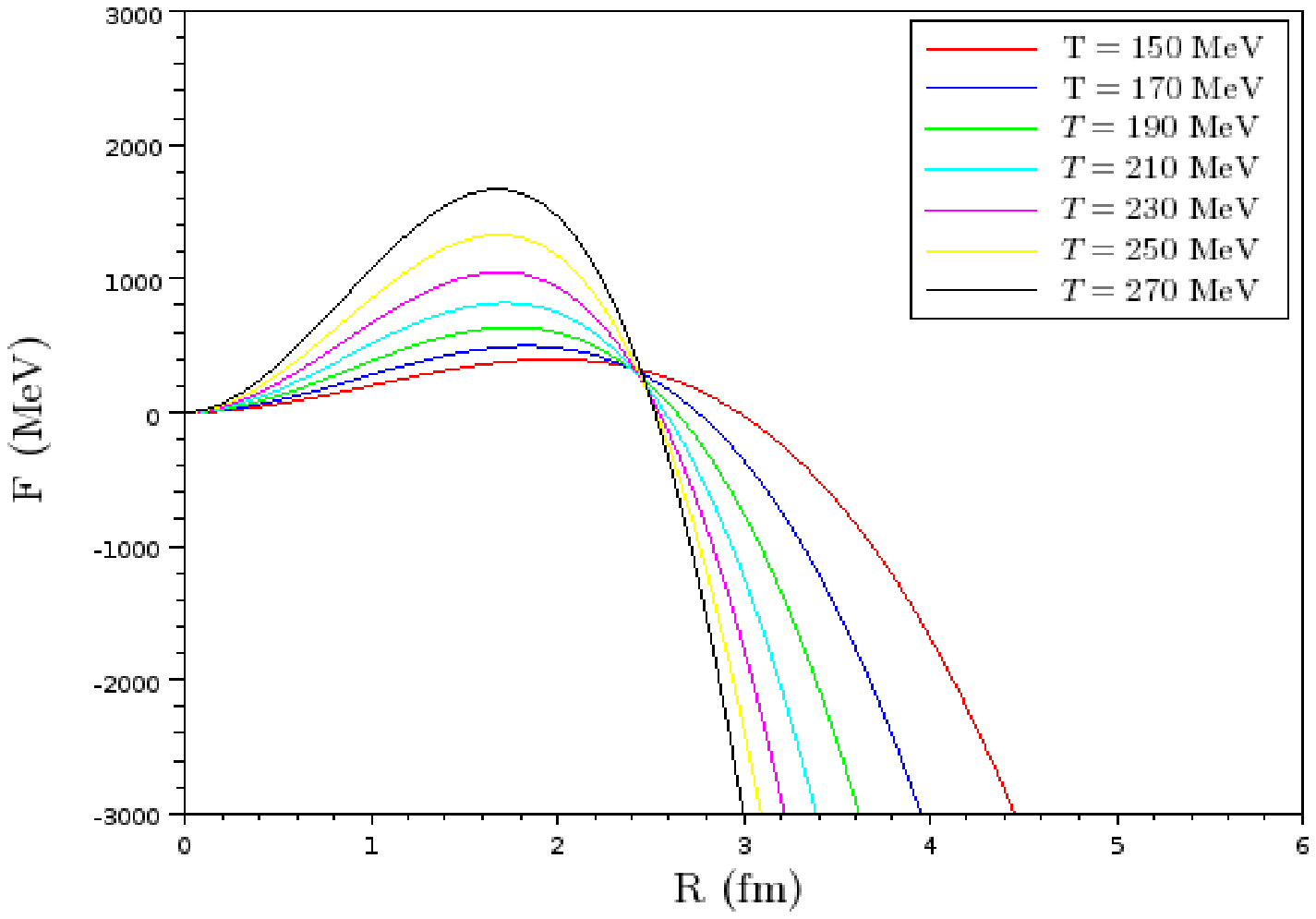}
\caption{ The free energy vs.~R~
at~$\gamma_{q}=1/14~$, $\gamma_{g}=64\gamma_{q}$ for the various values of temperature.}
\label{fig-3}       % Give a unique label
\end{figure}

\begin{figure}[htb]
\centering
%\sidecaption
\includegraphics[width=7cm,clip]{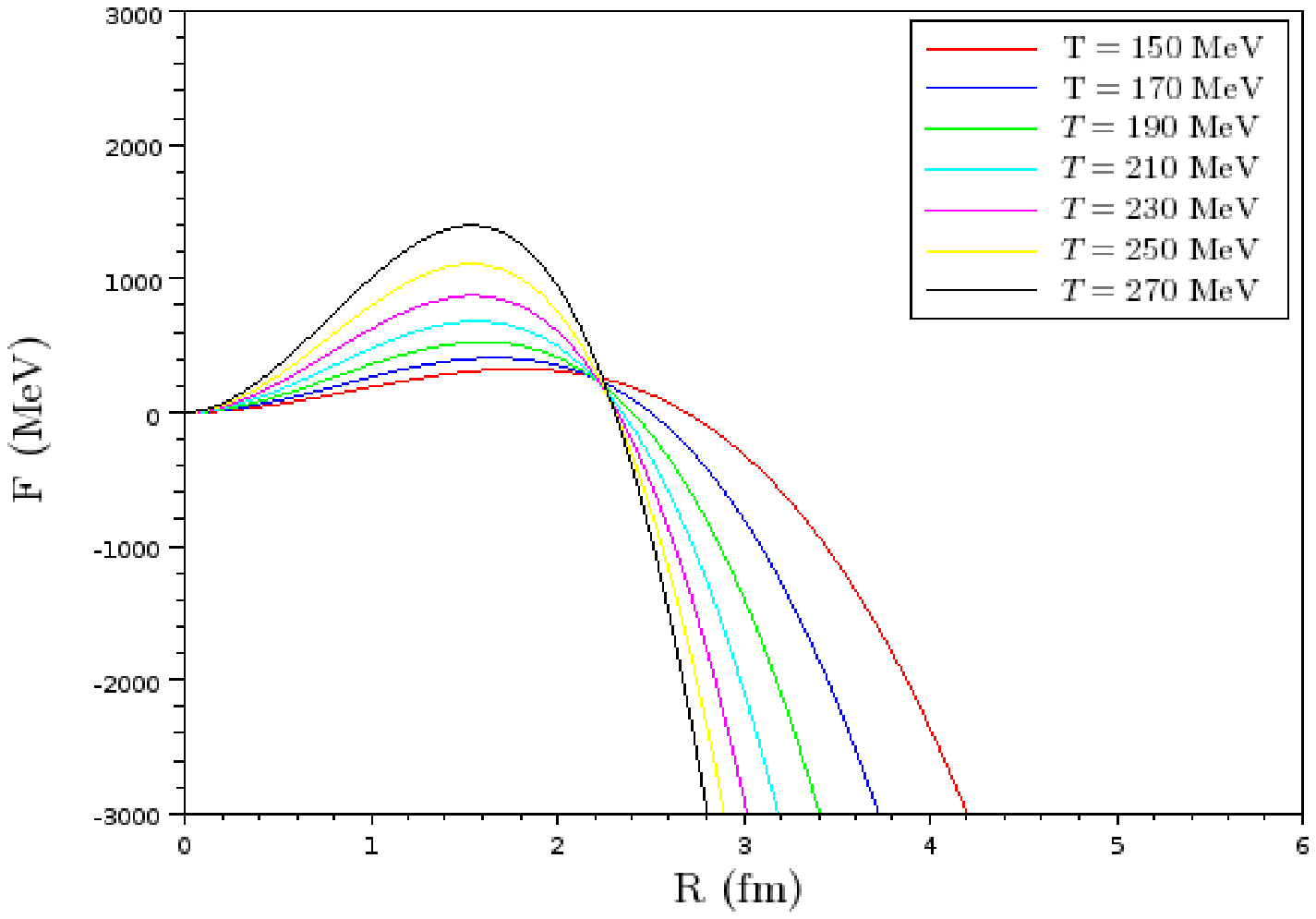}
\caption{The free energy vs.~R~
at~$\gamma_{q}=1/14~$, $\gamma_{g}=68\gamma_{q}$ for the various values of temperature.}
\label{fig-4}       % Give a unique label
\end{figure}

\begin{figure}[htb]
\centering
%\sidecaption
\includegraphics[width=7cm,clip]{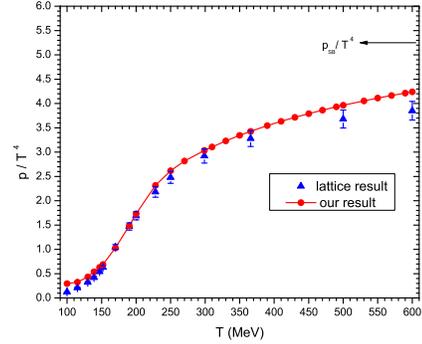}
\caption{Pressure vs.~T~
at~$\gamma_{q}=1/14~$, $\gamma_{g}=68\gamma_{q}$.}
\label{fig-5}       % Give a unique label
\end{figure}

\begin{figure}[htb]
\centering
%\sidecaption
\includegraphics[width=7cm,clip]{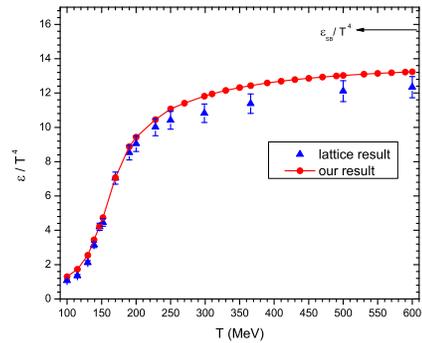}
\caption{Energy density vs.~T~
at~$\gamma_{q}=1/14~$, $\gamma_{g}=68\gamma_{q}$.}
\label{fig-6}       % Give a unique label
\end{figure}

\begin{figure}[htb]
\centering
%\sidecaption
\includegraphics[width=7cm,clip]{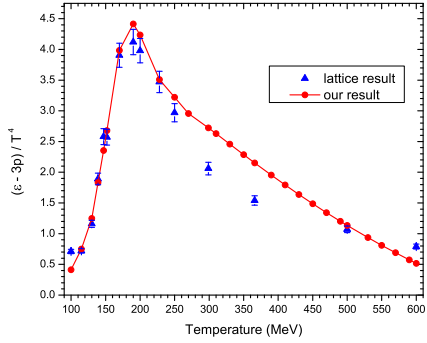}
	\caption{Interaction measurement vs.~T~
at~$\gamma_{q}=1/14~$, $\gamma_{g}=68\gamma_{q}$.}
\label{fig-7}       % Give a unique label
\end{figure}

\begin{figure}[htb]
\centering
%\sidecaption
\includegraphics[width=7cm,clip]{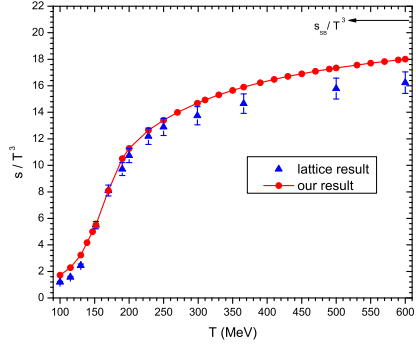}
\caption{Entropy vs.~T~
at~$\gamma_{q}=1/14~$, $\gamma_{g}=68\gamma_{q}$.}
\label{fig-8}       % Give a unique label
\end{figure}

\begin{figure}[htb]
\centering
%\sidecaption
\includegraphics[width=7cm,clip]{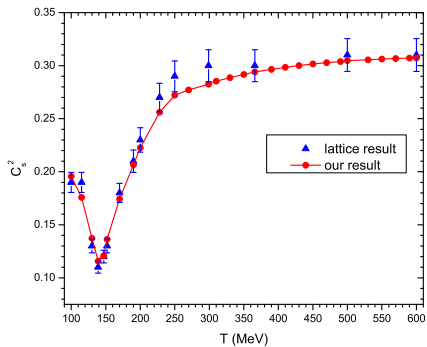}
\caption{Speed of sound vs.~T~
at~$\gamma_{q}=1/14~$, $\gamma_{g}=68\gamma_{q}$.}
\label{fig-8}       % Give a unique label
\end{figure}

\begin{figure}[htb]
\centering
%\sidecaption
\includegraphics[width=7cm,clip]{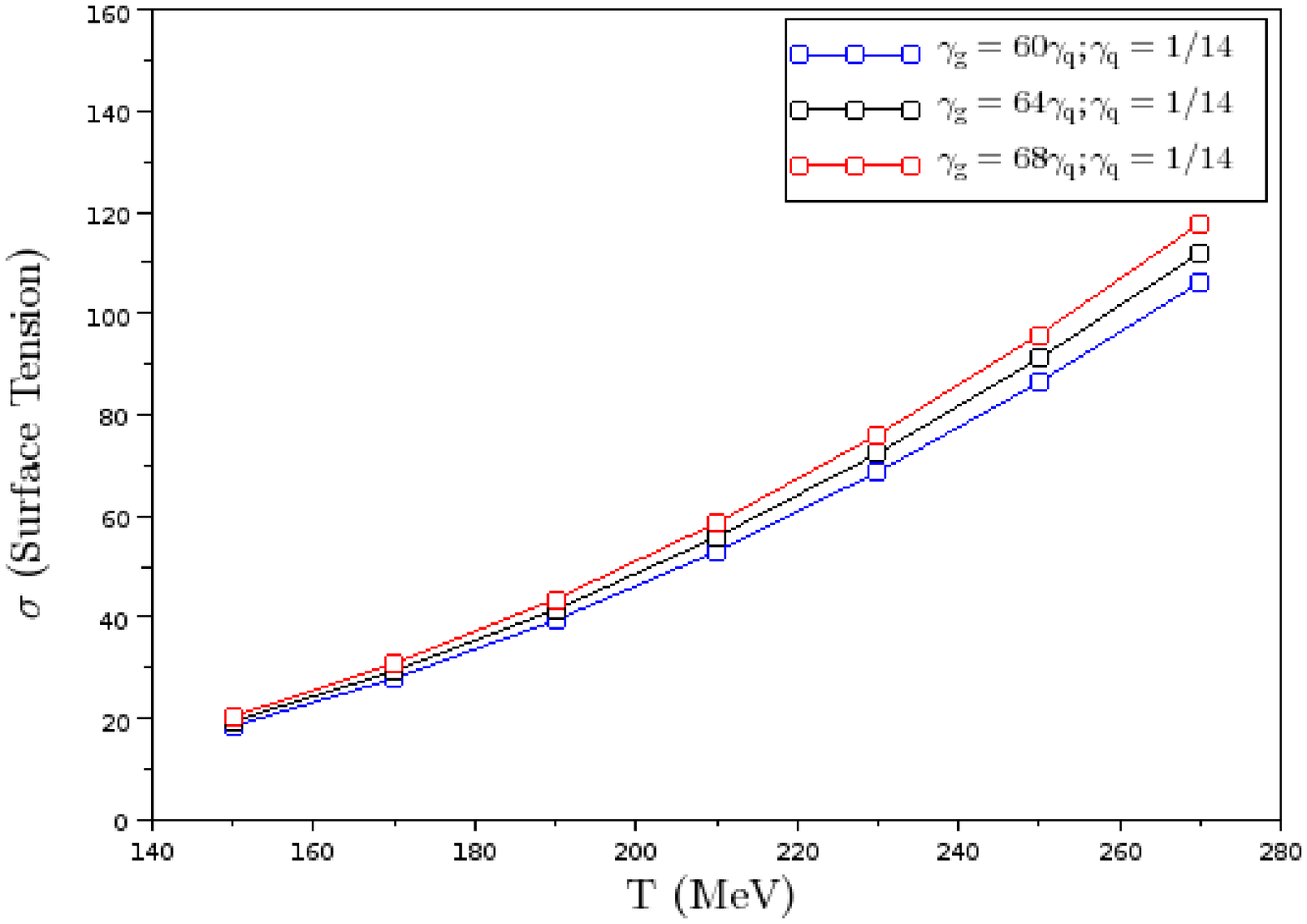}
\caption{The Surface Tension vs.~T~
at~$\gamma_{q}=1/14~$, different~ $\gamma_{g}$.}
\label{fig-8}       % Give a unique label
\end{figure}

\begin{figure}[htb]
\centering
%\sidecaption
\includegraphics[width=7cm,clip]{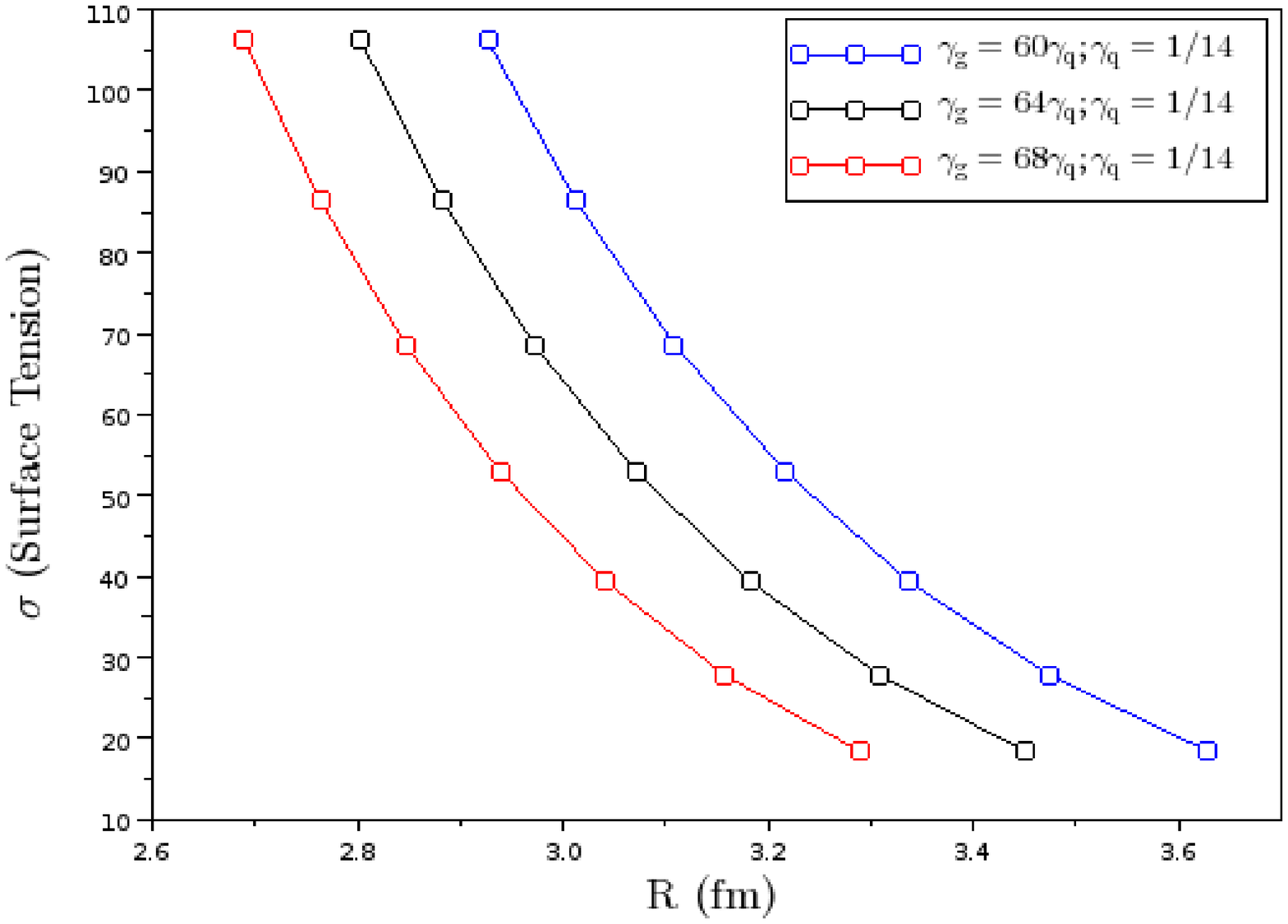}
\caption{The Surface Tension vs.~R~
at~$\gamma_{q}=1/14~$, different~$\gamma_{g}$.}
\label{fig-8}       % Give a unique label
\end{figure}

\begin{figure}[htb]
\centering
%\sidecaption
\includegraphics[width=7cm,clip]{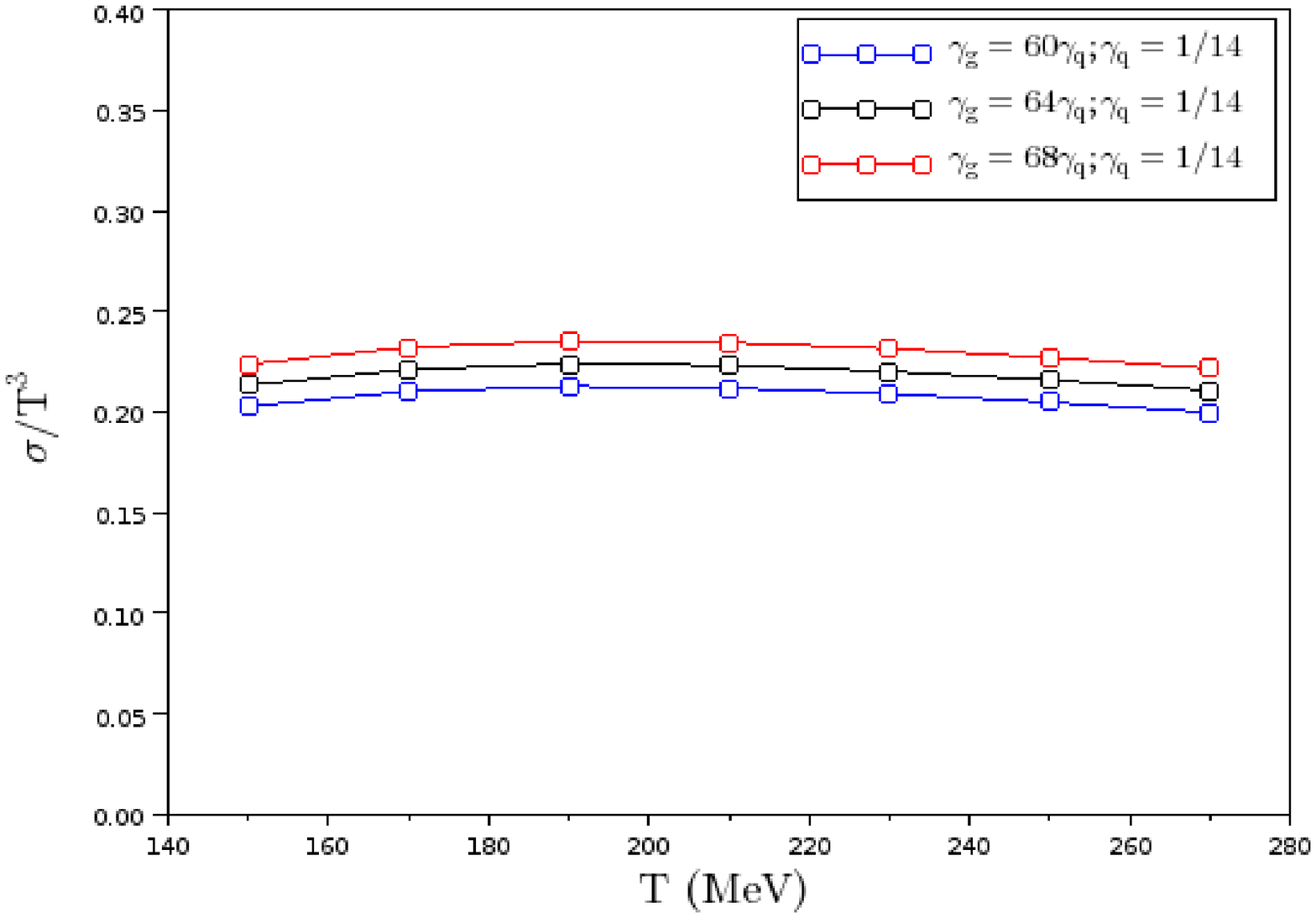}
\caption{The Surface Tension/$T_{c}^{3}$ vs.~T~
at~$\gamma_{q}=1/14~$, different~$\gamma_{g}$.}
\label{fig-8}       % Give a unique label
\end{figure}

\section{The free energy and Surface tension with loop Corrections}
The fermion and boson free energy after incorporating   
one to three loop corrections in the density of state is~\cite{mb}: 
\begin{equation}\label{3.20}
F_i = -\eta T g_i \int dk \rho_{q,g} (k) \ln (1 +\eta e^{-(\sqrt{m_{i}^2 + k^2}) /T})~,
\end{equation}
where $\eta=+ve$ gives the contribution of bosonic particle and 
$\eta=-ve$ gives the contribution
of the fermionic particles.
The extreme in the potential with minimal approximation is obtained 
by minimizing the confining potential in terms of momentum and it 
is approximately defined as:
\begin{equation}\label{3.19}
V(k_{min})=[\frac{8 a_{1}\gamma_{qg}N^{\frac{1}{3}} T^{2} \Lambda^4 }{27 \pi^{2}} ]^{1/6},
\end{equation}
where $N=(4/3 )[12 \pi / (33-2 n_{f})]$. Yet in this brief paper
the cut off potential
is taken differently from
the above value of minimum potential in order to get more
closer result to lattice. Due to the appropriate choice of this minimum
value we get more similar outputs to lattice with the larger contribution in 
the entire calculation of free energy of boson and fermion. 
It implies that the minimum cut off in
the integral tends to produce highly accurate in energy spectrum
of fermion and boson. It also helps in erasing
the infra-red divergence of the integration in evaluation with
the magnitude of $\Lambda$ and
$T$.
 $g_{i}$ is degeneracy factor (color and particle-antiparticle degeneracy)
which is $6$ for
quarks and~ $8$~ for gluons.
The inter-facial energy
obtained through Ramanathan et al's model~\cite{ramanathan,weyl} 
with a suitable parameter of
the hydrodynamic effects is also given as:
\begin{equation}\label{3.22}
  F_{interface}=\frac{1}{4}\gamma_{qg} R^{2}T^{3}. 
\end{equation}
The energy is used to replace the bag energy of MIT model. It gives minimum
shortfall in comparison to the energy used by MIT model. In addition
to these particles, we add some light particle 
hadrons and the corresponding energies are 
calculated using their masses~\cite{balian}

\begin{equation}\label{3.25}
F_{h} = (d_{i}T/2\pi^2 ) v \int_0^{\infty} k^2 dk \ln (1 - e^{-\sqrt{m_{h}^2 + k^2} / T}).
\end{equation}
where $d_{i}$ is the degeneracy factor for the different
light hadronic particles and $m_{h}$ is the light hadron corresponding masses.
Here we use only maximum light hadron as their involvement in the
reaction plane is more dominent during the time of collision. 
So total fermions involved in 
system are taken as:
dynamical quark masses $m_u = m_d = 0 ~ MeV$
and $m_s = 0.15 ~ GeV$. Therefore
the total free energy $F_{total}$ as a combination of fermion and boson is as:
\begin{equation}
 F_{total}=\sum_{j} F_{j}~+~F_{h},
\end{equation}
~ where $j$ stands for $u$,~$d$,~$s$~quark, interface and gluon. 

We further calculate
the surface tension of QGP fireball under the three loop correction
and as before, the evaluation is done by the 
difference of free energy between QGP phase and the light element 
hadrons phase. It implies that there is a sharp boundary between
hadron and quark phase, neglecting the short life mixed phase which
really balance the pressure and keep the system in
chemical equilibrium. Though it has been reported the importance
of the mixed phase in the calculation of surface tension due to 
large finite size and its effects disturbed on the surface 
tension~\cite{yasu,som}. So
the difference in energy of the two phases is defined after neglecting 
the finite size effects and shape contribution. It is therefore given as:
\begin{equation}
\Delta F =- \frac{4 \pi}{3}R^3 \Delta P + 4 \pi R^2 \sigma
\end{equation}
where the first term represents pressure difference between
QGP matter and hadronic matter. The second term represents
the contribution from the surface tension. The surface tension is evaluated
through the minimizing process of the above expression with respect to droplet 
size $R$.
So, the surface tension formula is obtained as:
\begin{equation}
Rc=\frac{2 \sigma}{\Delta P}~ or~ \sigma =\frac{3 \Delta F}{4 \pi R_{c}^2}
\end{equation} where, $\Delta F$ is the change in the free energy and $R_{c}$
is the corresponding critical radius obtained at the transition point
from quark droplet to hadron droplet.

\section{Results:} The free energy calculation is done through 
the inclusion of three loop to our earlier one loop 
correction model in the interacting mean-field potential. The calculated
outputs are represented by the corresponding figures. Due to the inclusion 
of loop corrections, 
the free energy evolution of QGP droplet changes a lot in terms of its
amplitude as well as the stability of droplet formation. 
The droplet sizes are found to be slightly changed with the
droplet size. This size is accordingly obtained with the parametrization
factor introduced in the model. 
The free energies are therefore shown in the figures 1,2 and 3 with the
different parameter factors and their stability also increases
with the change of gluon flow parameters. 
It indicates that the parameters really play a significant role to regularize 
the stability of the droplets and it acts as a magic number in modeling
the fluid dynamics. 
In Fig.$1$, we show the free energy of QGP
droplet at the particular
quark and gluon flow parametrization factor 
$~\gamma_{q}=1/14,~\gamma_{g}=60 \gamma_{q}$. Similarly, in Fig.$2$
it is free energy of QGP droplet with slightly lesser stability from the 
figure$1$. The droplet is obtained at the quark and gluon flow parameter
$~\gamma_{q}=1/14,~\gamma_{g}=64 \gamma_{q}$. Again we plot free energy
with increased gluon flow parameter. In Fig.$3$ the stability
of droplet is lesser, yet the droplet is obtained
at the parametrization factor $~\gamma_{q}=1/14,~\gamma_{g}=68 \gamma_{q}$. 
So we get a number of free energy evolution at different parameters showing
a slight difference in the stability of droplet.
Choosing one of the best stable droplet, then we try to picture the 
thermodynamic characteristics of the particular droplet at the particular
parameter value. These thermodynamics are presented by the equation of state
(EOS), pressure, interaction measurement, entropy,speed of sound and the 
surface tension with radius and temperature. These are shown in the figures.
 
In Fig.$4$ we see the pressure with the variation of temperature
and the plot is compared with the lattice data. The pressure
is almost similar with the lattice. In order to obtain a better
result much closer to the lattice data the minimum cut off
of the integration has been modified with the rescale momentum rather than
taking the exact minimum momentum cut off. If we put the the exact
minimum cut off then pressure at higher temperature does not change
and match with the lattice but at low tempearture it becomes very large.
Therefore, in order to keep the results closer to lattice from low to high
temperature a necessary modification has been introduced according to
momentum and temperature.
\par Similarly in Fig.$5$ we plot the energy density of our model and compare
with the lattice. Here the result is again similar in the entire range of 
temperature. This is also done with same scale in the integration
of energy density. It indicated that the model has a little drawback at the
lower temperature as the energy density is higher in comparison with lattice
if we look into our earlier cut off value. By changing the cut off
to scale of the lattice the model completely fit the the spectrum of 
the lattice trajectory.\\
Now by looking these above two plots in Fig.$6$, we plot
the interaction measurement with the temperature again. In the plot 
there is no much difference with the result produced by the 
lattice data. This indicates that the model can describe
the QCD phase structure in a very good way by modifying the cut off
in the integration of the free energy and other thermodynamic relations.
In the model we need  very small value of lower limit calculating
from theoretical expression and it then gives the different pictures in the
spectrum of QCD phase structure. In the lower value, it gives higher value
and as the limit is higher then it approximately falls to the lattice.
It is the reason that at higher temperature the result produced by
the present model lies within the range of the comparative results~\cite{dhar}.
\par Now we see the entropy picture and it is shown in Fig.$7$.
Entropy follows the same output of the lattice. This represents a good
presentation about the order of equilibrium. So the system has stable
flow in its movement from deconfined phase of QGP to confining phase of
hadrons. 
Again we plot the speed of sound and it is shown in Fig.$8$. Our data is 
compared with the lattice
data. The speed of sound absolutely follows the same curve traced by 
the lattice. It indicates that the speed of sound in our model results 
in producing almost the exact solution in the transformation of the QGP fluid 
movement producing the ideal liquid flow.
\par Now we look at the stability formation of the droplets.
We plot the different figures $9,10$ and $11$ of the surface tension. These
are the characteristic features to explain the surface tension.
 Fig.$9$ shows the variation of the surface tension with 
the size of droplet. As the size of droplet increases the surface tension
decreases and somewhere the droplet disappear and no surface tension 
is obtained. This can be shown that the surface 
tension tends to reach the horizontal axis representing the size of the droplet.
Again in Fig.$10$ we plot the surface tension with the variation of temperature.
As the temperature of the system increases the surface tension is found 
to be linearly increasing due to the formation of smaller size of the droplet.
At the larger temperature, it can be identified from the free energy formation
that the size of the droplet is decreased with the increasing temperature.
For the smaller temperature the size is large leading to
smaller value of the surface tension. This is the property of liquid 
drop formation. At last in Fig.$11$ we plot the the constancy of the surface 
tension as indicated by the other model that there is constancy if we 
look the plot of $\sigma/T^3$ with temperature $T$. The result is  
shown in our earlier paper also.
This indicates that the model is quite competent to explain the QCD phase
structure and other thermodynamic pictures after a simple modification
in the lower cut off of the integration of free energy with the scale of the
lattice to obtain more closer results with the 
lattice data~\cite{kaja,iwasaki,michael}
So the inclusion of three loop correction in the mean field potential
improve with increasing parameters
and enhance the stability of QGP droplet
with a slight modification in the cut off of the integration with
the scale of the larger momentum.
\par The parameter obtained with the three loop represents Reynold's 
like number showing the characteristic features of
fluid dynamics and these type of parameter can be
reproduced depending on the fluid. So in QGP fluid of early universe
expansion, we have such parameters to explain the equilibrium movement
of the QGP dynamics.
It means quark and gluon parameter may have some other 
characteristic features to determine the stability formation of QGP droplets.

\section{conclusion:}
The results show the formation of free energy and its improvement
by the presence of
three loop correction in the mean field potential. 
The effects on droplet formation is increased with the increase
in stability when the loop correction is increased. These results are indicated by the figures of 1 to 3 with the increase in the parametrization
values. As the loop correction is increased the free energy formation is also
increased depending on the number of particles involved in the system.
This has been observed from our earlier works that in one and two loop
correction we have obtained different size of the droplets depending on the
constituent particles. So the formation of stable droplet is effected by 
the parameter factor and particles. Taking into account of all these
factor, we attempt to produce more closer results to lattice 
and we produce very much similar results by approximating the number 
of particles and the lower cut off value of the momentum.
In addition to these results, we represents the thermodynamic properties
and surface tension to picture the QCD phase structure
and the stability of the droplets. The correction of three loop
really enhances more in all the thermodynamics relations in comparison to
two loop and one correction and it is more closer toward the lattice
results~[7,17]. It indicates that our simple model with three loop 
correction with the dynamical flow parameter
can produce better EOS of QCD and enhance the stable droplet formation
and thus the possible
solution is that the evolution of QGP fireball is more steady fluid dynamics 
depending on some kind of dynamical parameter
representing the characteristic features of forming the stable droplets.

\subsection{\bf Acknowledgments:}

We thank the retired Prof. R. Ramanathan for giving valuable time
in discussion about the manuscript's complete form.
The author, S. S. Singh thanks the university for providing research 
and development grants for this work successful.

\end{document}